\documentclass[12pt]{article}
\usepackage{graphicx} 
\usepackage{authblk}
\usepackage{hyperref}

\usepackage[margin=0.8in]{geometry}

\title{Whole-Genome Phenotype Prediction with Machine Learning: Open Problems in Bacterial Genomics}
\date{}

\usepackage{amsmath}
\usepackage{amssymb}
\usepackage{setspace}

\usepackage{lineno} 

\author[1]{Tamsin James}
\author[1]{Ben Williamson}
\author[1]{Prof. Peter Tino}
\author[1,2]{Dr. Nicole Wheeler}

\affil[1]{University of Birmingham, School of Computer Science, UK}
\affil[2]{University of Birmingham, Institute of Microbiology and Infection, UK \newline \href{mailto:txj287@student.bham.ac.uk}{txj287@student.bham.ac.uk}, \href{bsw978@student.bham.ac.uk}{bsw978@student.bham.ac.uk}, \href{n.wheeler.bham.ac.uk}{n.wheeler.bham.ac.uk}, \href{p.tin@bham.ac.uk}{p.tino@bham.ac.uk}}

\begin{document}

\maketitle


\abstract{How can we identify causal genetic mechanisms that govern bacterial traits? Initial efforts entrusting machine learning models to handle the task of predicting phenotype from genotype return high accuracy scores. However, attempts to extract any meaning from the predictive models are found to be corrupted by falsely identified ``causal'' features. Relying solely on pattern recognition and correlations is unreliable, significantly so in bacterial genomics settings where high-dimensionality and spurious associations are the norm. Though it is not yet clear whether we can overcome this hurdle, significant efforts are being made towards discovering potential high-risk bacterial genetic variants. In view of this, we set up open problems surrounding phenotype prediction from bacterial whole-genome datasets and extending those to learning causal effects, and discuss challenges that impact the reliability of a machine's decision-making when faced with datasets of this nature.}

\section{Introduction}

The goal of bacterial genome-wide association studies (bGWAS) is to identify genetic variants that influence a trait or phenotype (\cite{san2020current}). These studies traditionally employ statistical methods to perform population genomic analyses to yield a list of candidate genes or genetic markers associated with a phenotype, and have been a significant contributor in uncovering numerous genetic loci that are causally related to a phenotype, e.g., resistance to an antibiotic (\cite{collins2018phylogenetic, jaillard2018fast, lees2018pyseer, earle2016identifying, brynildsrud2016rapid}). Improvements in whole-genome sequencing techniques have led to the generation of increasing amounts of data, creating an impracticality surrounding functional investigations of all loci individually. However, this up-scaling has lead to the prediction of a greater number of significantly associated loci despite efforts to minimize false discovery rate.

Machine learning (ML) algorithms are an obvious successor to bGWAS that may more effectively find signal in genetic noise. To date, existing algorithms have been applied to the data with little to no adaptation (\cite{wang2022practical, pearcy2021genome, deelder2019machine, shi2019antimicrobial}). Researchers are finding that these ML models fail to reliably generalize to out-of-distribution examples (\cite{chalka2023advantage}, \cite{hu2024assessing}), and frequently identify false positive associations (\cite{pearcy2021genome}). In addition, they have found that removing all known causal variables from a model does not meaningfully impact model accuracy (\cite{nguyen2018developing}). Accordingly, it is crucial to understand the nuances of both genomic data and ML techniques to properly inform analyses, enabling them to mitigate effects that arise from model- and data-specific constraints, for integration with ML pipelines for causal variant detection. 

Predictive models do not inherently distinguish between causative and associative relationships. So we raise the question: what course of action should we take to develop a pipeline capable of distinguishing between causal and spurious features in bacterial genomic data? In this paper we present open problems in applying ML to phenotype prediction, with the goal of uncovering causal variants in bacterial datasets. We also discuss challenges specific to the intersection of causal ML and bacterial genomics, and use a dataset of 4,140 \textit{Staphylococcus aureus} isolates to illustrate these challenges with examples.

\section{Genotype-to-phenotype mapping} \label{Sec:GP_mapping}

We first set up the problem of linking genotype to phenotype under the assumption of fixed environmental conditions. 

Consider the sets of all biologically feasible genotypes and their corresponding phenotypes collected in the genotype and phenotype spaces $\mathcal{X}$ and $\mathcal{Y}$, respectively. The set $\mathcal{X}$ is finite (although vast due to the combinatorial nature of genetic variations), while $\mathcal{Y}$ can be finite or infinite (even a continuum). The sets represent an idealized, comprehensive view of all possible genetic and phenotypic variations within a bacterial population, reflecting both observed and unobserved possibilities. In nature, these spaces are constrained by biological and evolutionary limitations, with less fit genotypes being under-represented or absent. 


We assume the existence of a ground truth genotype-phenotype (GP) mapping function $\Theta: \mathcal{X} \rightarrow \mathcal{Y}$ assigning to each genotype $x \in \mathcal{X}$ the corresponding phenotype $y = \Theta(x) \in \mathcal{Y}$.

At a given time $t$, the potentially observable genotype and phenotype sets $\mathcal{X}_t \subseteq \mathcal{X}$ and $\mathcal{Y}_t \subseteq \mathcal{Y}$, respectively, are determined by the historical evolution process up to time $t$. In laboratory settings, the genotype space is further constrained by the experimental design, which can limit or direct the evolutionary process by controlling environmental conditions, selecting specific genotypes, or applying selective pressures.

One may collect an {\em empirical dataset} $\mathcal{D} = \{ (x_i, y_i)| x_i \in \mathcal{X}_t, y_i \in \mathcal{Y}_t, i=1,2,...,n\}$ of $n$ observed genotype-phenotype pairs by sampling from an unknown underlying distribution $P_t$ over $\mathcal{X}_t \times \mathcal{Y}_t$, using a sampling strategy that heavily biases the resultant set of data points. The distribution $P_t$ is dynamic, continuously evolving due to genetic drift, environmental pressures, and other evolutionary processes. Evolutionary processes lead to the formation of distinct genetic clusters, or strains, within the population. Different clusters exhibit significant divergence in genetic and phenotypic traits, especially in isolated environments, leading to $P_t$ with significant multi-modal structures. Note that due to possible "observational noise" the conditional distribution $P_t(y \vert x)$ over phenotypes $y \in \mathcal{Y}_t$, given a genotype $x \in \mathcal{X}_t$, may not be fully concentrated on the ground truth $\Theta(x)$. Observational noise may refer to variability in the phenotypes caused by measurement errors or epigenetic factors.

Our goal is to build a {\em predictive model} $f: \mathcal{X} \to \mathcal{Y}$ approximating $\Theta$ using the empirical dataset $\mathcal{D}_t \subset \mathcal{X}_t \times \mathcal{Y}_t$,  such that $f(x) \approx \Theta(x)$ for all $x \in \mathcal{X}$. The challenge is that we have at our disposal only a {\em limited sample} from the subset $\mathcal{X}_t \times \mathcal{Y}_t$ of the full space $\mathcal{X} \times \mathcal{Y}$ over which we would like to build the predictive model. An additional challenge is posed by the need to represent the input genotypes $x$ and output phenotypes $y$ in a form suitable for the machine learning technique employed. Assume that our parametrized machine learning predictive model (with parameters $w \in W$) takes as inputs and outputs elements of the input and output spaces $X$ and $Y$, respectively. The genotypes $x \in \mathcal{X}$ are represented through their feature representations $\phi(x) \in X$, $\phi: \mathcal{X} \to X$. Analogously, the phenotypes $y \in \mathcal{Y}$ are represented through $\psi: \mathcal{Y} \to Y$. The genotype feature representations $\phi(x)$ may not contain all the relevant information for the phenotype prediction. Due to these challenges, an additional set of potentially informative covariate variables $z \in Z_t$ is introduced to form extended inputs $(x,z)$ to the parametrized machine learning model $F: X \times Z \times W \to Y$. Examples of covariates may include the laboratory method used during data collection, or patient related factors. Hence, the phenotype prediction for a genotype $x \in \mathcal{X}$ with additional co-variates $z \in Z_t$ is realized through the parametrized machine learning model (with parameter values $w \in W$) as
\[
\hat y = \psi^{-1}(F(\phi(x),z;w)).
\]


\subsection{Representation spaces} \label{sec:representation_spaces}

To facilitate analysis and modeling, the actual genotype space $\mathcal{X}_t$ must first be mapped into a suitable representation space $\mathcal{R}$. This chosen representation space $\mathcal{R}$ can be any mathematical structure appropriate for representing genotype data, such as graphs, tensors, matrices or sequences. A marker space $\mathcal{M}$ serves as an intermediate representation that captures specific genetic markers derived from the raw genotype data that are locatable sequences on the genome (e.g., SNPs, k-mers, unitigs, or genes etc) through the processing of raw genetic data (i.e., sequencing or genotyping, alignment, variant calling, annotation and marker extraction).

We define a general embedding function $\rho: \mathcal{M} \rightarrow \mathcal{R}$ which maps the marker space $\mathcal{M}$ into a suitable representation space $\mathcal{R}$ of \textbf{embedded features} that facilitates analysis and modeling, allowing each input genotype $x \in \mathcal{X}_t$ to be represented as $\rho(x)$ in the new space $\mathcal{R}$, $\rho: \mathcal{X}_t \to \mathcal{R}$. The embedding function $\rho$ should preserve relevant genotypic properties.

Finally, the model representation space $X$ defines the space of \textbf{encoded features} prepared for input into the particular ML model. This is represented by a map $\tau: \mathcal{R} \to X$, giving us the genotype representation $\phi= \rho \circ \tau:\mathcal{X}_t \to X$ . As outlined above, the genotype features may be extended with additional informative co-variates $z \in Z$ which may capture nonlinear relationships and interactions that are not explicitly included in representations $\phi(x) \in X$.

We may summarize the composite function $\phi$ by a hierarchical mapping framework of representation spaces,

\[
\mathcal{X}_t \rightarrow \mathcal{M} \xrightarrow{\rho} \mathcal{R} \xrightarrow{\tau}  X.
\]


\subsection{Task 1}\label{sec:task1}

Given an extended training set $\tilde{ \mathcal{D}} = \{ (x_i, z_i, y_i)| \ x_i \in \mathcal{X}_t, z_i \in Z_t, y_i \in \mathcal{Y}_t, i=1,2,...,n\}$, the training process (adaptation of the parameters $w$) can be posed as an optimization problem. The task may be defined as \textbf{\textit{optimize the parameters $w$ by solving optimization problem}}:

\begin{equation}\label{eq:objective}
    w_{\text{opt}} = \arg\min_w \sum_{i=1}^n \ell(\psi^{-1}(F(\phi(x_i),z_i;w)),y_i),
\end{equation}

determined by an appropriate loss function $\ell: \mathcal{Y}_t \times \mathcal{Y}_t \rightarrow \mathbb{R}_{\geq0}$. 

\section{Genetic fine mapping}\label{Sec:fine_mapping}

In bacterial fine mapping, the goal is to identify \textbf{causal genetic variants} -- specific DNA sequence alterations within a population that directly influence phenotypic traits. This involves distinguishing true causal variants from those merely in linkage disequilibrium (LD) -- the phenomenon resulting in non-random association of alleles at different loci -- with causal ones and spuriously associated with the phenotype. In bacterial populations, LD extends throughout the entire genome due to clonal replication and leads to strong correlation structures across genome representations.


\subsection{Task 2}\label{sec:task2}

\textbf{\textit{Identify a subset of genomic markers $\mathcal{M}^* \subseteq \mathcal{M}$ that are the true causal variants influencing the phenotype.}}

Given that the learning of the GP mapping is well-posed, the goal is to optimally disentangle causal from non-causal features such that the true causal variants may be identified. Since the GP task on its own may not be specific enough to pinpoint the causal features, additional measures to constrain the feature space may need to be deployed. For example, \cite{aliee2023conditionally} suggest a deep neural network with architecture and loss functions specifically informed by causality analysis.

Besides direct task-driven architectural  and loss function manipulation, additional regularization pressure $R_{causal}$ may be introduced (e.g. in an additive form) to help zoom on the causal features. For example, denoting the genotype encoding mapping using marker set
$\mathcal{N} \subseteq \mathcal{M}$ by $\phi(x; \mathcal{N})$, the size of the marker set by $\vert \mathcal{N}\vert$, and a class of ML models input-compatible with this encoding by 
$\mathcal{F}_{\phi(\cdot; \mathcal{N})}$, we can express the overall optimization problem as:


\begin{equation}
\label{eq:fine-mapping}
\mathcal{M}^* = \arg\min_{\vert \mathcal{N} \vert; \mathcal{N} \subseteq \mathcal{M}} 
 \ \Bigl[ \min_{F \in \mathcal{F}_{\phi(\cdot; \mathcal{N})}}
\sum_{i=1}^n \ell(\psi^{-1}(F(\phi(x_i; \mathcal{N}),z_i)),y_i) + \lambda R_{causal}(F) \Bigr].
\end{equation}

\section{Open problems}\label{Sec:open_problems}

Given the optimization problem \ref{eq:objective}, the goal is to find the parameter set  $w_{\text{opt}}$ that minimizes the total loss over the training data. The function $F(\phi(x_i),z_i;w))$ produces some intermediate output based on the transformed input $\phi(x_i)$, covariates $z_i$, and the parameters $w$, in the phenotype representation space $Y$. The inverse mapping $\psi^{-1}$ is then applied to this output to obtain a prediction $\hat{y}_i$ before computing the loss with respect to the true target $y_i$.  In some cases the map $\psi$ can be simply be the identity map.

The core issue arises because the mapping from the parameter space to the space of predictive models $F$ may not be injective. This means that this mapping is many-to-one and the desired parameter setting is not uniquely identifiable.
However, even if the mapping from the parameter space to the space of predictive models was injective, because our training sample is finite and sparse, several parameter settings $w$ can lead to exactly (or almost exactly) the same phenotype predictions on the sample genotypes.

Hence, multiple parameter settings $w \in W$ can produce the same predictions $\hat{y}_i \in \mathcal{Y}_t$ on the training data. Consequently, \textbf{the inverse mapping from predicted phenotypes back to parameters is one-to-many}: different parameter settings $w$ can lead to the same predicted phenotype $\hat{y}_i$ for given inputs $\phi(x_i)$ and $z_i$.

As a result, the model cannot uniquely determine the parameters $w$ that map the input genotypes $\phi(x_i)$ to the observed phenotypes $y_i$. This non-uniqueness manifests as multiple solutions to the optimization problem, where different parameter sets minimize the loss function equally well, rendering the problem ill-posed. This can have serious consequences for approaches that detect ``important" input features (e.g. causal variants) based on the learned parameter vector $w$ (e.g. Automatic Relevance Determination or Matrix Relevance Learning (\cite{Wipf2007ANV}, \cite{5430874}). 


\subsection{Open problem 1}\label{sec:OP1}

\textbf{OP1.A - }\textbf{\textit{Is it possible to reformulate the genotype-to-phenotype mapping task to be well-posed?}}

Learning of the genotype-to-phenotype mapping is inherently ill-posed (when unconstrained), so in order to render the task as well-posed, we need to address the violations of Hadamard's criteria -- existence, uniqueness, and stability -- by introducing additional constraints and modifications to the problem structure.

Given that there exists a ``ground truth" formulation of the GP mapping that is unknown to us, we further ask:

\textbf{OP1.B - }\textbf{\textit{What are the necessary conditions and representations required to achieve a well-posed task?}}

To make learning of the GP mapping well-posed, additional constraints must be employed in the formulation of data representations (mappings $\phi$ and $\psi$) and of the predictive model $F(\cdot; w)$ itself. These aspects can involve e.g. feature selection, dimensionality reduction techniques, and incorporation of domain knowledge. 
One may attempt to enforce the constraints, for example, through additive regularization
\begin{equation}
\label{eq:OP1-objective}
\min_{F \in \mathcal{F}_{\text{constrained}}} \sum_{i=1}^n \ell(\psi^{-1}(F(\phi(x_i),z_i; w)),y_i) + \lambda R(F),
\end{equation}
of the core optimization problem \eqref{eq:objective}. Here $\mathcal{F}_{\text{constrained}}$ is a restricted function space incorporating prior knowledge and potentially of reduced complexity compared to $\mathcal{F}$, $\lambda \geq 0$ is a regularization parameter that controls the trade-off between fitting the data and satisfying the regularization term, and $R(\cdot)$ is a regularization functional that imposes additional constraints on $F$.


\subsection{Open problem 2}\label{sec:OP2}

\textbf{\textit{Does there exist a well-posed genotype-to-phenotype mapping that also satisfies requirements for bacterial fine mapping? And if so, what are the necessary conditions and representations to achieve this?}}



\subsection{Challenges in genotype-to-phenotype prediction}

Assessing the well-posedness of predicting bacterial phenotypes from genotypes involves examining whether the optimization problem defined in Equation \ref{eq:objective} meets Hadamard's criteria. According to Hadamard's criteria, a problem is well-posed if a solution \textbf{exists}, is \textbf{unique}, and depends continuously on the input data (is \textbf{stable}) (\cite{hadamard2014lectures}). 

In the context of bacterial genotype-to-phenotype mapping,  these criteria are often unmet, rendering the problem \textbf{ill-posed}. Genome-wide LD is the primary factor contributing to non-injectivity in the mapping. Additionally, non-injectivity can arise from several other sources, including limited dataset sampling, loss of information in feature representations, unmeasured confounders and observational noise, as well as the complexities of the model and constraints within the function space.

To elucidate these challenges, we examine an example real dataset: the \textit{Staphylococcus aureus} pangenome, representing the empirical dataset $\mathcal{D}_t$. This dataset encompasses genetic variation across the entire genome $\mathcal{X}_t$ and associated phenotypes $\mathcal{Y}_t$, specifically binary measures of antibiotic resistance across 16 different drugs. While our analysis focuses on this collection of isolates, the identified challenges are broadly applicable to various bacterial species, albeit to differing extents. Additional analyses have been previously detailed (\cite{wheeler2019contrasting}), providing a foundation for our discussion.



\section{Limited sampling in the dataset}

The empirical dataset $\mathcal{D}_t$, from which we construct the extended training set $\tilde{\mathcal{D}}$, is a limited and potentially biased sample from the underlying distribution $P_t$ over $\mathcal{X}_t \times \mathcal{Y}_t$, and may not contain sufficient information to uniquely determine the optimal parameters $w_{\text{opt}}$. Due to evolutionary processes like genetic drift and selection, $P_t$ can be multi-modal and may not capture the full variability of genotype-phenotype relationships. 

The dataset may lack diversity, missing critical genotype variations that influence the phenotype, leading to an underdetermined system where multiple parameter settings fit the available data. 

Without adjusting for confounders, a solution $w_{\text{opt}}$ may exist since there are parameters $w$ that minimize the loss function $\ell$ over the training set $\tilde{ \mathcal{D}}$, but it may not represent the true underlying relationship. Additionally, the existence of a reliable solution may also be prevented due to insufficient data and small sample size, due to the dataset $\mathcal{D}_t$ only representing a \textit{limited sample} from the subset $\mathcal{X}_t \times \mathcal{Y}_t$. Data scarcity is a prevalent issue in bacterial genomics, with whole genome sequencing dataset sizes typically on the order of $10^2 - 10^3$ isolates for most species (\cite{mosquera2023genome}). The \textit{Staphylococcus aureus} dataset contains $n=4140$ samples.




\section{Information loss in representations}

Information loss in the feature representations $\phi: \mathcal{X} \rightarrow X$ for genotypes and $\psi: \mathcal{Y} \rightarrow Y$ for phenotypes leads to non-injectivity in the mapping $F$ in optimization \ref{eq:objective} by failing to preserve all relevant details necessary for accurate prediction. 

The mapping $\phi: \mathcal{X} \rightarrow X$ may fail to preserve all relevant genetic information (i.e., due to dimensionality reduction or feature selection) necessary for accurate prediction.

Similarly, $\psi: \mathcal{Y} \rightarrow Y$ may simplify phenotypic information and obscure differences between phenotypes that are important for learning the mapping. Phenotype space compression -- converting continuous or multiclass traits into binary categories -- focuses the learning task by reducing the complexity of the output space, but at the expense of information loss. 

Following the hierarchical framework in Section \ref{sec:representation_spaces}, the \textbf{forward mappings} $\rho$ and $\tau$ must prevent the loss of variant information and preserve unique genetic information relevant to causality. Ideally, they should maintain a one-to-one mapping for features between spaces to ensure full identifiability for fine-mapping tasks, or should be informed with relevant biological properties to enhance interpretability when extracting causal insights. 

Practical examples of information loss arise when applying dimensionality reduction and feature selection techniques to manage high-dimensional genetic data. While these methods are essential for computational efficiency and mitigating overfitting, they can inadvertently eliminate relevant genetic factors crucial for accurately modeling the true mapping function $F$ in the optimization task. Excluding such factors risks producing suboptimal models that fail to capture the underlying biological relationships.

\textbf{Feature selection} methods reduce dimensionality of the feature space, and by extension constrain the function space $\mathcal{F}$, by selecting a subset of features before modeling (or as part of the model fitting) to focus the model on the most informative genetic variants. Given the dataset $\mathcal{D}_t$ and representation function $\phi: \mathcal{X}_t \to X=\mathbb{R}^D$, the feature selection objective is to define a selection function $\sigma: X \rightarrow X_S$, where $X_S = \mathbb{R}^d \subset X$ is the reduced representation space with $d<D$
selected features $S \subseteq \{1,2,\dots,D\}$. When using a genotype matrix $M$, $\sigma$ maps each genotype representation $\phi(x_i) \in \mathbb{R}^D$ to a reduced $(1\times d)$ feature vector $\sigma(\phi(x_i)) \in \mathbb{R}^d$. Commonly used techniques include performing statistical tests (using univariate tests to select genetic features, e.g., SNPs, unitigs etc, significantly associated with the phenotype), recursive feature elimination (iteratively removing features based on their importance to the model), and embedded methods (algorithms that perform feature selection as part of the model training process).

\textbf{Dimensionality reduction} is the process of transforming data from a high-dimensional representation space $X$ to a lower-dimensional space $X' \subset X$ while preserving essential properties or structures of the original data, useful for the phenotype prediction task at hand. 

\subsection{Genetic markers}

The choice of marker space $\mathcal{M}$ critically shapes the representation genotype space $X$, influencing the well-posedness of the GP mapping by enhancing biological relevance. For example, LD-induced redundancy may be mitigated through marker selection, with unitigs and genes supporting more robust GP mapping and precise fine mapping, but at the risk of excluding causal variants.


A single nucleotide polymorphism, or \textbf{SNP}, is a single base-pair change at a specific genomic location between isolates, hence they capture single base-pair variations but miss complex genetic changes. In practice, insertions and deletions (indels) may also be included in this feature representation to account for these more intricate genetic alterations.

\textbf{k-mers} represent contiguous nucleotide sequences of fixed length $k$ extracted from raw genomic data, offering broader genomic variation coverage but resulting in high-dimensional, redundant feature spaces, and facilitate the assembly of genomes via construction of a \textbf{de Bruijn Graph (DBG)} -- a graph structure where each node represents a k-mer, and edges connect nodes that overlap by $k-1$ bases. 

\textbf{Unitigs} and \textbf{compacted DBG (cDBG)} structures aggregate overlapping k-mers (non-branching paths within a DBG) into unique, contiguous sequences, providing more efficient and less redundant representations. 

\textbf{Genes} consolidate multiple variants, lowering dimensionality and improving interpretability, though they sacrifice resolution and exclude non-coding regions. 


\textbf{Uniqueness} in the GP mapping is not guaranteed due to the \textbf{high dimensionality} and redundancy in genomic data (high \textbf{multicollinearity} due to LD); multiple parameter sets may yield the same minimal loss. The genome $x \in \mathcal{X}$ is typically measured in base pairs, with size falling anywhere within an approximate range of 100 kbp to 10 Mbp for bacteria (\cite{rodriguez2022genomic}). The \textit{Staphylococcus aureus} genome is typically $\sim$2.8Mbp (\cite{holden2004complete}), but the dataset grows with the capture of genetic variants. The genome representations $\phi(x) \in X_t$ are typically high-dimensional. For example, the \textit{Staphylococcus aureus} dataset is noted to follow a unitig representation that results in $p\approx 1.2$m parameters, notably reduced in comparison to alternative variant representations (i.e., k-mers). 

Extending the task of constructing a well-posed GP problem to bacterial fine mapping (Section \ref{sec:task2}) -- inferring causal relationships from observational data -- poses additional significant challenges, especially given the lack of controlled interventions that are typically available in experimental settings. Namely, the \textbf{marker space} $\mathcal{M}$ must include all potential causal variants with precise, ideally unique, genomic locations to ensure \textbf{identifiability} (uniquely associating genetic variants $\mathcal{M}^* \subseteq \mathcal{M}$ with phenotypic variations) and \textbf{interpretability} (biologically contextualizing relationships between genetic variants and phenotypic traits).

\subsection{Representing the genotype space} \label{sec:g_matrix}


In practice, one considers a finite set of markers $\mathcal{M} = \{m_1, m_2, ..., m_p\}$, for example the set of unitigs -- nodes of a cDBG $G_t$. The representation space $\mathcal{R}$ is then the space of all finite length paths through the graph $G_t$. Hence, each genotype $x \in \mathcal{X}_t$ is represented by the corresponding path $\rho(x)$ in $G_t$. For our machine learning model, the paths in $G_t$ (elements of $\mathcal{R}$) can be encoded e.g. by marking out the unitigs visited by the path $\rho(x)$. In other words, the representations space is the $p$-dimensional binary hypercube, $X = \{0,1\}^p$, where the $j$-th coordinate of $\phi(x) = \tau(\rho(x)) \in X$, $\phi(x)_j$, is 1 if and only if the unitig $m_j$ is visited by the path $\rho(x)$ (i.e. if the unitig is contained in the genotype $x$), and 0 otherwise.

The genotypes of the data set {$\mathcal{D} = \{ (x_i, y_i)| x_i \in \mathcal{X}_t, y_i \in \mathcal{Y}_t, i=1,2,...,n\}$} can then be compactly represented via a \textbf{genotype matrix} $M \in \{ 0,1 \}^{n \times p}$,

\begin{equation*}
    M = 
\begin{pmatrix}
    \phi(x_1)_1 & \phi(x_1)_2 & \cdots & \phi(x_1)_p \\
    \phi(x_2)_1 & \phi(x_2)_2 & \cdots & \phi(x_2)_p \\
    \vdots & \vdots & \ddots & \vdots \\
    \phi(x_n)_1 & \phi(x_n)_2 & \cdots & \phi(x_n)_p 
\end{pmatrix}
\end{equation*}

Each row $\phi(x) = (\phi(x_1)_1, \phi(x_1)_2, \cdots,  \phi(x_1)_p)$ corresponds to the genotype of sample $i$ across all $p$ genetic features.

Of course, other types of representations are possible and in general
entries $\phi(x_i)_j$ can be binary values (presence or absence of a feature), categorical variables, or numerical values, depending on the nature of the genetic data representation. If a suitable dimensionality reduction technique is applied, the final genotype representations may be of dimensionality $d<p$.

While the genotype matrix is widely used (\cite{ma2020increased}, \cite{macesic2020predicting}, \cite{burgaya2023bacterial}), it is important to consider whether alternative representations might better capture the biological complexity of the data for certain modeling approaches, such as graph-based (\cite{yang2021end}) or sequence-based representations (\cite{wiatrak2024sequence}).



\section{Unmeasured confounders and observational noise}




\subsection{Population structure}

Population structure arises from the evolutionary history and genetic relatedness within the sample population. It affects both genotype and phenotype distributions in $P_t$, introducing correlations that can confound the genotype-phenotype relationship (spurious associations between genotypes and phenotypes). \textbf{It is a hard requirement that confounders must be accounted for either implicitly and/or explicitly for a GP mapping problem to be well-posed.} 

\textbf{Explicit modeling} involves directly incorporating known confounders into the ML model through additional features or adjustments. Extending to fine mapping, reliable causal inference demands \textbf{causal sufficiency} -- no unmeasured confounders influencing both the cause and effect variables -- which necessitates explicit modeling of confounders. Explicit modeling approaches rely on accurately approximating the unmeasured confounders through various techniques. Population structure may be handled explicitly via additional covariates $z$ or separately via regularization, typically through extracting principal components (PCs) from genotype data using methods such as Principal Component Analysis (PCA) (\cite{kavvas2020biochemically, kouchaki2019application}), or regularization with a similarity matrix $K$ (\cite{farhat2019gwas, yurtseven2023machine}).



Derived from the genotype data, $K$ captures how genetically similar each pair of individuals is within a population, which can be calculated using measures such as genetic distances, phylogenetic distances from reconstructed trees, or other appropriate similarity metrics. The matrix $K$ influences the model through regularization, penalizing large differences in predicted phenotypes $\hat{y}$ for individuals that are genetically similar. An example (\cite{yurtseven2023machine}) of enforcing this additional constraint is given by:

\begin{equation}
\min_{F \in \mathcal{F}_{\text{constrained}}} \sum_{i=1}^n \ell(\psi^{-1}(F(\phi(x_i),z_i; w)),y_i) + 
\lambda {\bar F}^\intercal K {\bar F}.
\end{equation}
Here, $F: X \times Z \times W \to \mathbb{R}$ and 
\[
{\bar F} = ( 
F(\phi(x_1),z_1; w),
F(\phi(x_2),z_2; w),
...,
F(\phi(x_n),z_n; w)
)^\intercal \]
is the vector of model predictions across the data sample.

\textbf{Implicit modeling} relies on the ML model's inherent ability to learn and adjust for confounders through complex pattern recognition without directly incorporating specific confounder variables into the model. Complex ML models, such as deep neural networks, inherently possess the capacity to learn and adjust for confounders through enhanced nonlinearities and richness in the functions, provided sufficiently large data sample is available. However, bacterial datasets typically exhibit limited sample sizes, constraining the applicability of these complex models that demand extensive amounts of data to achieve robust generalization.


Consequently, the prevalent method for implicitly accounting for population structure in bacterial studies is the use of of cross-validation techniques, which are well-suited for simpler models, enabling identification of scenarios in which the model has over-fit. These methods are focused on model evaluation and generalizability, via assessing how well a model handles confounders. Traditional methods using random partitioning (\cite{pincus2020genome}) can inadvertently place related strains in both sets, thus inflating performance metrics due to the model effectively memorizing strain-specific features rather than learning generalizable patterns. Strain-specific cross-validation partitions the data based on genetic clusters (\cite{lees2020improved}). In this approach, entire strains are held out during each fold of training and validation, which tests the model’s ability to generalize across different genetic backgrounds and reduce the risk of overfitting to specific strains.



\subsection{Observational noise}
Observational noise refers to random variability in the data that cannot be attributed to the variables being studied, making it challenging for models to learn accurate mappings.

Observational noise may be attributed to measurement errors (e.g., sequencing errors and phenotype measurements), stochastic gene expression (e.g., random fluctuations in antibiotic resistance levels unrelated to genetic variability), processing errors (e.g., DBG assembly), or technical/detection limitations (e.g., low expression genes falling below detection thresholds).


\section{Model complexity and function space limitations}




The complexity of the predictive model $F$ and and limitations of the function space it operates within can cause non-injectivity in the mapping from parameters $w$ to predicted phenotypes $\hat{y}$. An overparameterized model, with a high-dimensional parameter space $W$, may contain redundancies or non-identifiable parameters. Conversely, an underparameterized model may lack the capacity to approximate the true genotype-to-phenotype mapping $\Theta$. 
 
Limited sample sizes inherent to bacterial datasets impose significant constraints on the choice of predictive models, often being forced to rely on simpler models, which may risk being underparameterized. To mitigate the lack of inherent ability in capturing complex biological relationships, it is essential to enhance these simpler models with additional constraints or incorporate prior biological knowledge.


\subsection{ML view of datasets}\label{sec:ML_view_data}

Before adding additional constraints to address ill-posedness of the GP mapping task, it is essential to understand how ML models inherently ``see" datasets and how their assumptions might align (or conflict) with knowledge of bacterial population behavior. 

ML models typically assume that data points are \textbf{independent and identically distributed (iid)}, which is often violated in bacterial datasets due to evolutionary processes that introduce dependencies among genotypes -- genotypes present within the actual distribution over $\mathcal{X}_t \times \mathcal{Y}_t$, from which $\mathcal{D}_t$ is sampled, cannot be assumed to be independently generated.

The \textbf{stationarity} of the data distribution is another critical assumption, yet bacterial populations may experience shifts over time and geographical location driven by selective pressures and founder effects, leading to non-stationary data generating processes. 

Furthermore, ML models assume that the dataset contains a \textbf{complete and representative set of features} relevant to the target outcome. However, the complexity of bacterial genomes and technological limitations can result in incomplete feature sets, where not all relevant genetic variants are observed. \textbf{Measurement accuracy} is also presumed, but in practice, sequencing errors and phenotype measurement inaccuracies introduce noise into the data. 

Domain knowledge is not inherently reflected within many ML models due to \textbf{uniform treatment of features} (all features being treated equally). The biological relevance of certain genetic markers may therefore not be captured within the model, and will limit the ability of a model to make reliable predictions by exclusion of these parameters.

Many ML models also assume \textbf{smoothness} and \textbf{stability} in the map on representations from genotypes to phenotypes $F: X \rightarrow Y$. Smoothness here refers to $F$ being continuous between metric spaces $(X,d_X)$, and $(Y,d_Y)$, where $d_X$ and $d_Y$ correspond to distance metrics that measure distances between elements $\phi(x_i) \in X$ and $\psi(y_i) \in Y$, respectively. $F$ is continuous if $\forall \epsilon > 0, \: \exists \delta>0$ such that $d_X(\phi(x_1), \phi(x_2)) < \delta \Longrightarrow d_Y(\psi(y_1), \psi(y_2)) < \epsilon$, where $\epsilon$ and $\delta$ are positive real numbers representing \textit{small} distances in the representation spaces $X$ and $Y$, respectively. However, in bacterial GP mappings, the ground truth $\Theta$ is often non-smooth due to causal sparsity, where minor genetic differences can lead to significant phenotypic changes. Predictive models assuming smoothness in $F$ may struggle to generalize when attempting to approximate $\Theta$ with a smooth mapping $f$.

\begin{figure}
    \centering
    \includegraphics[scale=0.5]{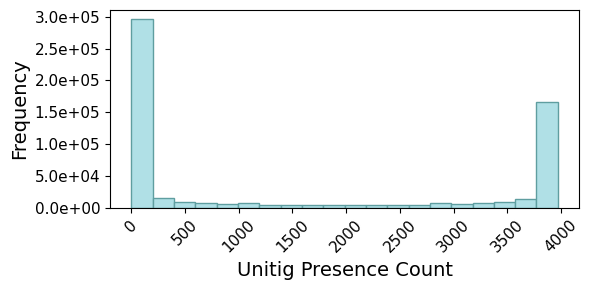}
    \caption{Histogram of unitig presence count across 4140 \textit{Staphylococcus aureus} isolates.}
    \label{fig:unitig_freq}
\end{figure}

For \textit{Staphylococcus aureus}, just 61 acquired genes and 82 mutations form the set of known causal mechanisms across 16 phenotypes (\cite{wheeler2019contrasting}). There is also additional sparsity in the feature set, where the majority of unitigs are only present in a handful of samples. Figure \ref{fig:unitig_freq} displays the occurrence rate for each of the 1,238,055 unitigs across all samples, with their frequency ranging from 1 to 4,139 out of 4,140 samples. A large portion of unitigs are only present in 83 (2\%) or fewer samples. This level of sparsity in the data exacerbates the issue of overfitting, making it much harder to generalize to new data.

\subsection{Choice of predictive model}\label{sec:model_choice}

This section focuses on models facilitated by a genotype matrix representation $M$ -- the predominant data representation for GP prediction tasks -- and does not cover models that inherently require alternative data representations

\textbf{Linear models.} Linear models assume a linear relationship between genetic features and the phenotype, wherein the genetic features influence the phenotype independently and additively. They provide a straightforward approach and interpretable results, with the addition of penalty terms better managing high-dimensionality (\cite{burgaya2023bacterial}, \cite{saber2020benchmarking}). This limits the function space $\mathcal{F}$ to linear functions with parameters $w^\intercal = (w_{\phi}^\intercal, w_z^\intercal$) -- where in the optimization \ref{eq:objective}, $F(\phi(x_i),z_i; w)) = w_{\phi}^\intercal\phi(x_i) + w_{z}^\intercal z_i$ represents a linear function of the transformed genotype $\phi(x_i)$ -- which are unable to capture the complex nonlinear interactions often present in bacterial populations. 

\textbf{Nonlinear models.} The relationship between features and the phenotype may be nonlinear and involve interactions. Expanding the function space $\mathcal{F}$ to include nonlinear functions increases flexibility but also the risk of overfitting if not properly regularized. Among the non-linear models, \textbf{tree-based models} have enjoyed particular attention (\cite{karanth2022exploring}, \cite{allen2021forest}, \cite{batisti2024optimising}). Decision trees and ensemble methods (random forests, gradient boosting) can naturally model interactions and nonlinearities. The phenotype can be predicted based on hierarchical decision rules derived from the genetic features. The function space $\mathcal{F}$ is constrained by the structure of the trees and hyperparameters like depth and the number of trees. Interpretability can sometimes be challenging due to the complexity of the ensemble models.

Parameter fitting can be performed in various settings, but if $F$ is formulated as a probabilistic model (conditional on an extended input $(\phi(x),z)$ a full distribution over phenotypes is provided), a \textbf{Bayesian framework} can be adopted. Bayesian approaches incorporate prior distributions over model parameters, allowing for natural integration of prior knowledge and uncertainty quantification. The posterior distribution over the model parameters $w$ is given by $p(w|\mathcal{D}) \propto p(\mathcal{D}|w)p(w)$, where $p(w)$ is the prior and the likelihood $p(\mathcal{D}|w)$ is provided by our probabilistic model.

\subsection{Feature attribution in fine mapping}

Interpretability is crucial for fine mapping to accurately distinguish causal variants from confounded associations, making feature attribution methods a popular choice for extracting causal insights. Ensemble methods such as random forests and gradient boosting machines, along with ability to handle high-dimensional data, provide insights into feature importances (\cite{buckley2021lessons}, \cite{rahman2018machine}, \cite{wassan2018paam}). 

\textbf{Feature importances} provide a quantitative measure of how much each genetic feature $\phi(x)_j$ contributes to the model's prediction $\hat{y}$, derived from $X$. A high feature importance score suggests that a particular feature has a strong association with the phenotype of interest.  

\textbf{SHAP values} estimate the contribution of each $\phi(x_i)_j$ to the model's prediction $\hat{y}_i$ for a given sample $i$, so can highlight specific genetic variants that consistently influence the phenotype across different strains or isolates. However, these methods capture \textit{associations} rather than \textit{direct causal relationships} (\cite{ma2020predictive}), therefore will be prone to highlighting regions of the genome that are spuriously associated with the phenotype rather than causally related, due to the high rates of LD in bacterial populations.


\section{Incorporating domain knowledge}



To address the challenges posed by the one-to-many inverse mapping from phenotypes to model parameters, integrating domain-specific knowledge into the genotype-to-phenotype mapping can be used to constrain the function space and improve model interpretability.


\subsection{Spatial dependencies}

Spatial dependencies refer to the relationships and interactions between genetic variants based on their physical locations on the genome, and possibly their positions within representational structures such as cDBGs by extension. Spatial dependencies can be integrated into the feature representation $\phi(x)$ by encoding the physical or representational positions of genetic variants. This integration allows the predictive model $F$ to account for the localized interactions and collective effects of adjacent genetic variants. For instance, variants that are physically proximal on the chromosome may be more likely to interact epistatically, influencing the phenotype in a manner that is not apparent when considering each variant independently. Spatial dependencies facilitate the incorporation of genomic domain knowledge, such as the tendency of causal variants to cluster within specific genomic regions or regulatory elements, and can constrain the model by prioritizing interactions among nearby variants 

\begin{figure}
    \centering
    \includegraphics[scale=0.4]{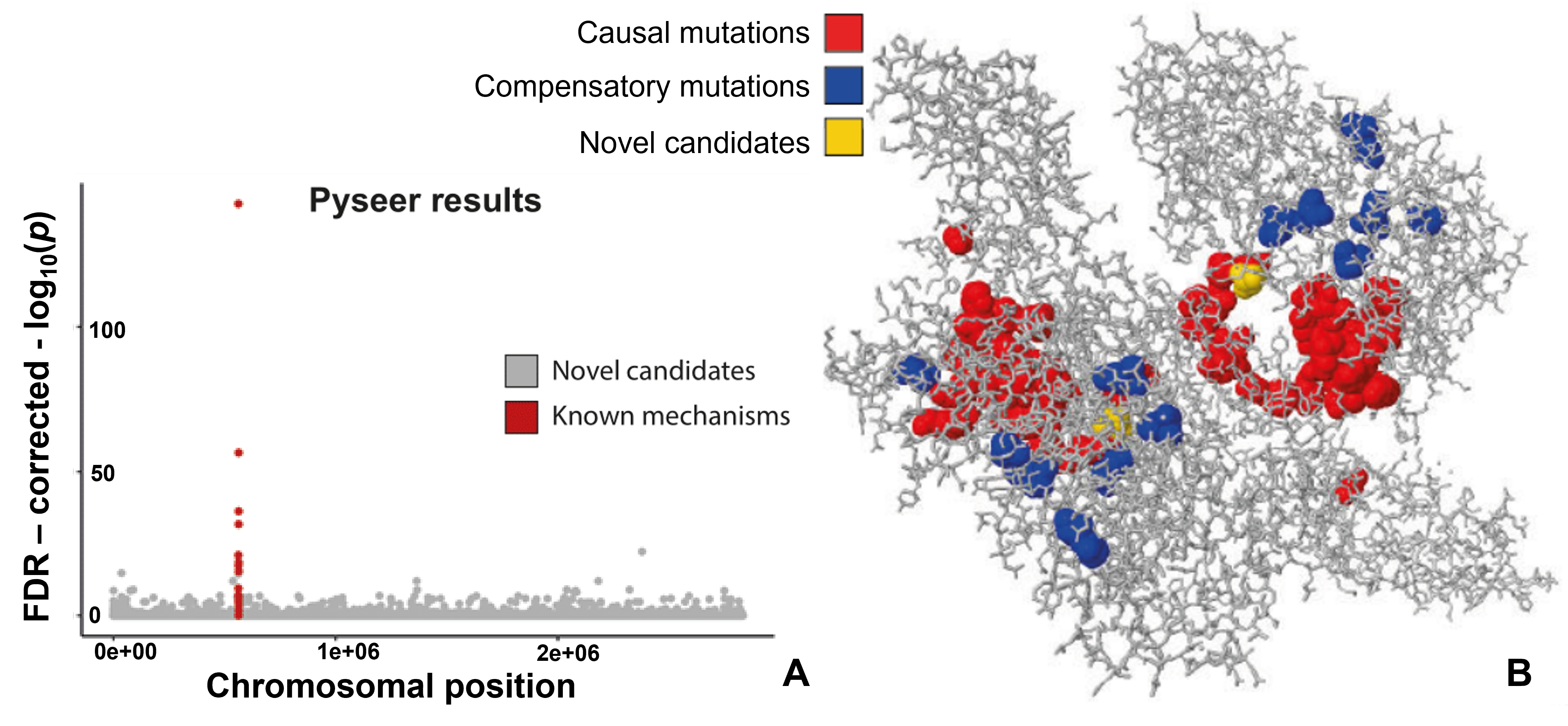}
    \caption{A: Manhattan plot for bGWAS single nucleotide polymorphism Pyseer results. B: 3-dimensional structure obtained experimentally of a single \textit{Staphylococcus aureus} protein. Figure adapted from \cite{wheeler2019contrasting}.}
    \label{fig:3d_structure}
\end{figure}

Figure \ref{fig:3d_structure}A illustrates a Manhattan plot for bGWAS results using Pyseer (\cite{lees2018pyseer}), highlighting the close proximity of causal variants in the genome. Figure \ref{fig:3d_structure}B depicts a 3-dimensional protein structure of the main chromosomal locus for fusidic acid resistance in \textit{Staphylococcus aureus}. It is evident that known causal mutations, along with compensatory mutations (mutations which co-occur with causal mutations to mitigate any negative impacts they have on protein function) and novel bGWAS candidates, exist within spatial ``hotspots'' -- novel candidates obtained via bGWAS (yellow) are close to known causative mutations (red). Most classical ML models accept data in a tabular format, where features are treated as independent variables and lack any representation of spatial dependencies (standard representation of genomic data discussed in Section \ref{sec:g_matrix}). 

\subsection{Prior causal knowledge}

In the formal framework, prior causal knowledge can be incorporated by selectively augmenting the feature representation $\phi(x)$ with information about known causal variants or by structuring the predictive function $F$ to prioritize these variants during the learning process. 



\section{Conclusion} 

Several challenges remain in applying machine learning to phenotype prediction in bacterial genomics. The primary issue is determining whether the genotype-to-phenotype mapping can be reformulated as a well-posed problem, including identifying the necessary conditions and representations for this transformation. If a well-posed mapping is achievable, the next challenge is ensuring it meets the specific requirements for bacterial fine mapping.

Key areas to address include developing robust representation spaces that preserve biological properties and causal information while minimizing spurious associations. This development depends on model choice and the effective integration of domain knowledge throughout the pipeline. Additionally, balancing model complexity with interpretability is essential for bacterial fine mapping. While more complex models may better capture biological mechanisms, they can reduce interpretability and require larger sample sizes than typically available. Therefore, interpretability must be maintained through complementary approaches, such as post hoc analysis or specialized interpretability techniques, to ensure biological relevance.



\bibliographystyle{acm}

\bibliography{reference}

\end{document}